\documentclass[pra,floatfix,preprint,nofootinbib,eqsecnum,12pt]{revtex4}
\usepackage{epsfig}
\usepackage{bm}% bold math
\usepackage{amsmath}
\input epsf
\numberwithin{equation}{section}

\def\Hc{{\cal H}}
\def\HP{$(H,|\Psi\rangle)$}

\def\igs{IGUSes}
\def\ig{IGUS}

\def\be{\begin{equation}}
\def\ee{\end{equation}}

\begin{document}
\vspace{1cm}

\title{Why Our Universe Is Comprehensible\footnote{Based in part on a talk given at the FQXI meeting in Banff, Canada, August, 2016}}
\author{James B.~Hartle}

\email{hartle@physics.ucsb.edu}

\affiliation{Santa Fe Institute, Santa Fe, NM 87501}
\affiliation{Department of Physics, University of California,Santa Barbara, CA 93106-9530}

\date{\today }

\begin{abstract}
Einstein wrote memorably that `The eternally incomprehensible thing about the world is its comprehensibility.' This paper argues that the universe must be comprehensible at some level for information gathering and utilizing subsystems such as human observers  to evolve and function. 
\end{abstract}

\maketitle

\bibliographystyle{unsrt}

%%%%111111111111111111111111111111111111111111111111111111111111111111111111111111

\section{Introduction}
\label{intro}

Albert Einstein wrote famously \cite{Ein36}:  
\begin{quote}\it The eternally incomprehensible thing about the world is its comprehensibility
\end{quote}
The thesis of this essay is that the world must be comprehensible in order for information gathering and utilizing systems (IGUSes)  including human beings like us to exist\footnote{ A more general discussion of IGUses will be given in Section \ref{iguses}. For more discussion in the context of complex adaptive systems see \cite{Gel94}. }. 

The argument   in a nutshell is that IGUSes exploit the regularities of the world in order to satisfy evolutionary imperatives like ``get food yes'', ``be food no'', ``make more yes'', etc.  A frog catching a fly is making use of a crude,  approximate, (and probably hard wired) form of Newton's laws of motion  summarizing regularities of the quasiclassical realm of our quantum universe. To contain IGUSes the universe must have regularities that can be exploited by IGUSes.
 
We comprehend our universe by discovering and understanding its regularities. A universe with IGUSes must therefore be comprehensible at least at the level necessary for IGUSes to function\footnote{It's clear from the text in \cite{Ein36} immediately before the quote above that for Einstein the comprehensibility  at issue includes everyday human experience, and not merely regimes that are foreign to us such as the interior of the Sun, or the very early universe.}.

This essay will amplify on these thoughts in the context of the decoherent (or consistent) histories quantum theory  of closed systems (DH) --- most generally the universe\footnote{For classic expositions, some by founders of the subject, see \cite{classicDH}. For a tutorial see \cite{Har93a}. }.  The purpose of this essay is not to understand exactly what Einstein meant by his aphorism. The author is not a historian and would not presume to comment on that.
But Einstein's statement is a natural starting point for a discussion of aspects of DH
in which we give more definite meanings to  terms like `world' and `comprehensible'. The following development is offered in that spirit. 

The comprehensibility of the universe would indeed be a mystery if, as observers of the universe, we were somehow outside it. That would be the case in in the Copenhagen formulations of quantum theory. Why should a physical system that we observe from the outside have any regularities at all? But we are not outside the universe. We are physical systems within the universe, subject to the laws of quantum mechanics but playing no preferred role in its formulation (cf \cite{Eve57}).  We are just special kinds of small fluctuations that have  evolved  over a very long time in a very large universe. The universe must exhibit regularities for these physical systems  to exist,  function, and proliferate. 

DH offers not just one way of comprehending the universe, but many, through different decoherent sets of alternative coarse-grained histories --- realms for short. The sets will differ in the variables that are followed by the histories. There can be descriptions that are complementary  somewhat like the position and momentum of a particle. Generally they will be at different levels of coarse-graining.
The universe is thus not comprehensible or incomprehensible. It is rather comprehensible in different ways at different levels of precision.  In the rest of the paper we will begin to address the question of which realms exhibit IGUSes.

%The preceding discussion explains why the universe is comprehensible at some level if it contains IGUSes.  But this simple result invites the question of what regularities lead to IGUSes.  In subsequent sections we will address this question in the context of a model quantum universe.

The paper is organized as follows. Section \ref{model} introduces a model quantum universe in a box to focus subsequent discussion. Section \ref{qcrealms} discusses the quasiclassical realms of everyday experience. Familiar IGUSes are described by histories of a quasiclassical realm. Section \ref{iguses} discusses \igs\ and their requirements. Section \ref{beyondqc} raises, but does not answer,  the question of whether the universe might exhibit realms  incompatible with  our quasiclassical ones with essentially different notions of \igs.  Section \ref{conclusion} contains summary conclusions.

 \section{A Model Quantum Universe}
\label{model}

\begin{figure}[t]
\includegraphics[width=3in]{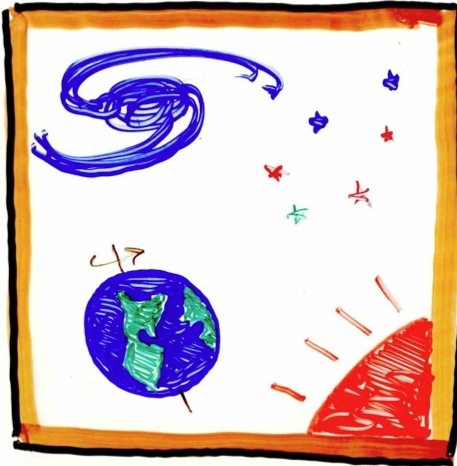}\hfill
\caption{A simple model of a closed quantum system is a universe of quantum matter fields inside a large closed box (say, 20,000 Mpc on a side) with fixed flat spacetime inside. Everything is a physical system inside the box --- galaxies, stars, planets, human beings, observers and observed, measured and measuring. The most general objectives for prediction are the probabilities of the individual members of decoherent sets of alternative coarse grained histories that describe what goes on in the box. That includes histories describing any measurements that take place there. There is no observation or other intervention from outside.  }
\label{box}
\end{figure}

To keep the discussion manageable, we consider a model closed quantum system in the approximation that gross quantum fluctuations in the geometry of spacetime
can be neglected. The closed system can then be thought of as a large (say
$\gtrsim$ 20,000 Mpc), perhaps expanding, box of particles and fields in a 
fixed, flat, background spacetime (Figure \ref{box}). There is thus a well defined notion of time in any particular Lorentz frame. The familiar apparatus of textbook quantum mechanics then applies ---  a Hilbert space $\Hc$, operators, states,  and their unitary evolution in time.  We assume a quantum field theory in the flat space for dynamics. Assuming a fixed spacetime means neglecting quantum gravity. 

The important thing is that everything is contained within the box --- galaxies, planets, observers and
observed, measured subsystems, and any apparatus that measures
them.  This is a model cosmology and the most general physical context for prediction. In this model Einstein's `world' is this box. 

%%%%%%%%%%%%%%%%%%%%%%%%%%%%%%%%%%%%%%%%%%%%%%%%%%%%%
%%%%%%%%%%%%%%%%%%%%%%%%%%%%%%%%%%%%%%%%%%%%%%%%%%%%%%%

\subsection{Decohering Alternative Histories of the Universe}
\label{histories}

The basic theoretical inputs for predicting what goes on in the box  are the  Hamiltonian $H$ governing evolution and quantum state  $|\Psi\rangle$, written here in the Heisenberg picture for convenience, and assumed pure for simplicity. Input theory is  then \HP.

The basic output of the theory are the probabilities of the individual members of sets of alternative coarse-grained histories of the contents of the box. A history is a sequence of events,  for example the history of the formation of the solar system, or the  events in human history.  A history is {\it coarse-grained} when it follows some variables and ignores the rest.  For example, a coarse grained history of the Moon's orbit might follow only the center of mass position of the Moon and ignore the positions of other particles in the Moon, and follow the center of mass position not at all times, but only at a sequence of times, and these positions only to an accuracy sufficient for comparing with  observation.  A history that also describes the Moon's rotation would be finer-grained. 

There must be negligible quantum interference between the individual coarse-grained histories in a set for its members to be assigned probabilities consistent with the usual rules of probability theory. Such sets are said to {\it decohere}.  A decoherent set of alternative coarse-grained histories of what goes on in the box is called a {\it realm}. 

Some coarse graining is necessary for decoherence. Ignoring some particles in the box so they constitute an environment for the rest of the particles  is widely used example (e.g. \cite{JZ85}).  A completely fine-grained set of histories of all the particles in the box would not decohere.  In quantum mechanics some information must be ignored to have any information at all.

More explicit and mathematical detail on how this all works is provided in the Appendix.

\section{Quasiclassical Realms and their Regularities}
\label{qcrealms}

The most striking observable feature of our indeterministic quantum universe is the wide range of time, place, and scale on which the deterministic laws of classical physics hold to an excellent approximation. 
Regularities that characterize classical physics are described by the familiar classical equations for particles, bulk matter, and fields, together with  the Einstein equation  governing the regularities of  classical spacetime geometry. 
Our observations of the universe suggest that a domain of classical predictability extends from a short time after the big bang to the far future and over the whole of the visible spatial volume. Were we to set out on a journey to arrive in the far future  at a distant galaxy we would count on the regularities of classical physics holding there just as they do here. 

A quasiclassical realm is a decoherent set of alternative histories of the box that is coarse-grained by familiar quasiclassical variables.  To use our model box universe as an example,  first imagine dividing the space inside the box into small volumes.  For each volume give the averages over  it of energy, momentum, and  the number of relevant conserved particles.   These are the usual hydrodynamic variables that occur, for example,  in the Navier-Stokes equation.  Specifying such spatial configurations at a series of times defines a set of alternative quasiclassical histories. Larger volumes define coarser-grained quasiclassical histories,  small volumes a finer-grained description.   The volumes must be large enough for the set of histories to decohere. Then the set is a quasiclassical realm\footnote{For a review of  quasiclassical realms with more references, see, eg.  \cite{Har08}.}.

Useful quasiclassical realms span an enormous range of coarse grainings. Quasiclassical realms describing everyday experience may be so highly coarse-grained as to refer merely to the features of a local environment.  At the other end of the range are the quasiclassical realms that are as fine-grained as possible given decoherence and quasiclassicality. Those realms extend over the wide range of time, place, scale and epoch mentioned above. 

By  predicting probabilities for the histories of a quasiclassical realm the theory \HP\ predicts probabilities for the evolution of localized systems  such as galaxies, stars, planets, biota, humans, tables, chairs, bacteria, etc., and most importantly for this paper ---  IGUSes.  

\subsection{Regularities of the Quasiclassical Realm}
\label{regularities}
A quasiclassical realm exhibits a regularity when the theory \HP\ predicts  high quantum probabilities for histories of quasiclassical alternatives that exhibit reliable  correlations in time or space or both.   The dynamical laws of classical physics define regularities in time.  Newton's laws of motion, Maxwell's equations,  the Navier-Stokes equation, the Einstein equation, and the laws of thermodynamics are all familiar examples. These classical equations are approximate consequences of the underlying quantum theory \HP\ (e.g. \cite{GH07}). They hold only in limited regimes of the universe and can be interrupted by quantum events. 

Regularities in space can arise  from symmetries in \HP.  The approximate homogeneity and isotropy of the present universe on scales above several hundred megaparsecs is an example. Regularities like the second law of thermodynamics arise from a quantum state in which the entropy associated with a coarse graining by quasiclassical variables was low all across the universe at an early time (e.g. \cite{GH07}).    

IGUSes  use approximations to the regularities of the laws of classical physics as in the example of a frog catching a fly.
But the frog is also using regularities of flies,  plausibly something like:  all flies are food,  big flies are better than small flies, flies tend to be near ponds, etc, etc. Such regularities typically arise from frozen accidents.

A frozen accident is a chance event whose consequences proliferated. For instance the chance events of mutation, recombination, and genetic drift produce the evolution of new species that proliferate across planets. The similarities between fly species arise from their common evolutionary origin; their differences arise from the frozen accidents of biological evolution.

The  regularities following directly from \HP\ can be expressed mathematically The regularities of specific systems like flies are generally not easily mathematically expressed. The frog may be said to be using some approximation to Newton's laws of motion, but the regularities of flies are of a different character.  They do not follow with high probability from \HP. They arise from chance accidents.  Individual  flies are, of course, subject to the general laws of classical physics as parts of a quasiclassical realm.  

IGUSes exploit the regularities that the universe presents to evolve and proliferate. Were there no regularities  to exploit there would be no IGUSes. 

\subsection{Coarse Graining and Comprehensibility}
\label{cg-comp}

We comprehend the universe by understanding its regularities in the context of specific realms like the quasiclassical ones.  This is how we interpret `comprehensible' in Einstein's aphorism. 
As mentioned in the Introduction there are different kinds of comprehension in different variables at different levels of coarse graining.  Following finer grained histories generally gives more comprehension. 
 We comprehend the orbit of the Moon by following its center of mass.  Comprehending the libration of the Moon requires a finer grained set of histories that also follows the Earth. 

Comprehension is limited by decoherence. For instance, we can never have a completely fine-grained understanding of the universe because the necessary set of histories would not decohere (e.g \cite{GH11}).

\section{IGUSes}
\label{iguses}                                                                                                                              

\subsection{What is an  \ig?}
\label{characterization}

As human observers of the universe both individually and collectively we are examples of information gathering and utilizing systems (IGUSes).  IGUSs are approximately localized subsystems of the universe characterized by the following three properties:
\begin{itemize} 
\item{\igs\ acquire information about their environment. }
\item{\igs\ use the regularities in the acquired information to update a model or `schema' of their environment and beyond.}
\item{\igs\  act on the predictions of this schema, typically acquiring new information in the process.}
\end{itemize}

There are many examples of IGUSes in our quasiclassical realm.  Individual humans are \igs, and so are societies of human beings. The humans pursuing science today constitute the human scientific \ig. Cockroaches and bacteria qualify as \igs\,  as would any system that would be called `living'.   A self-driving car is an IGUS, so is an airplane on autopilot, and a thermostat. Etc.,etc. 

All these examples of IGUSes can be seen as features of our quasiclassical realm. That is, they and their actions  can be described by particular histories coarse-grained by quasiclassical variables as described above\footnote{The observers  in Copenhagen quantum mechanics might be seen examples of \igs. But those observers were features of a  classical world separate from the quantum one.  In DH  \igs\ are subject to the quantum mechanical laws and describable in quantum mechanical terms.} .                                                             

\subsection{Regularity, Randomness, and Complexity}
\label{reg-rand}
Why do quasiclassical realms exhibit \igs?  One answer lies with what an \ig\ needs to function.   An IGUS requires reliable regularities to build its schema and function effectively.   But it also requires some randomness to evolve.  On Earth IGUSes  might have started when random fluctuation in circumstances produced the right chemical environment for life. The collective  IGUS that is a species also needs the randomness of mutation, genetic drift, or recombination to evolve. Beyond regularity and randomness the varied world of IGUSes  that we see today requires something like a fitness landscape that provide various niches for IGUSes to evolve into.  That is, they require some level of complexity.   Quasiclassical realms can provide the right mixture of regularity to exploit, randomness to evolve, and complexity for a fitness landscape  that leads to the presence of \igs.  In Section \ref{beyondqc} we raise the question as to whether there are other kinds of realm with \igs.

\subsection{\igs\ in the Expanding Universe} 
\label{cosmo}
To better understand the requirements for \igs\ let's think a little outside our model box and consider \igs\ in the expanding universe, modeled by letting the  box of Section \ref{model} expand. 

The evidence of the observations is that our universe expanded from a hot big  bang. In a simplified discussion we can usefully identity four subsequent epochs at different times  $t$ after the beginning at $t=0$.  

\begin{itemize}

\item{\it The Epoch of Quantum Gravity:}  A short era $t\lesssim 10^{-43} s$ in which the geometry of spacetime and matter fields exhibited large quantum fluctuations.  The histories constituting a quasiclassical realm are based on classical  and therefore start only after this \cite{Har08}.

\item{\it The Early Universe:}  The early universe consisted of a hot plasma of nucleons, electrons, and photons. As revealed by the observations of the cosmic background radiation (CMB) in this epoch the universe was homogeneous, isotropic and featureless to an excellent approximation (deviations of $1$ pt in $10^5$).  

\item{\it The Middle Universe:} As the universe expands the temperature drops. When it has dropped sufficiently,  electrons, protons and nucleons recombine and nuclei are synthesized. The initial  fluctuations from the quantum gravity era grow under the action of gravitational attraction. Eventually they produce the universe of galaxies, stars, planets, biota, and \igs\ that we find today.

\item{\it The Late Universe:}  The cosmological constant causes the universe to expand and cool exponentially quickly. Stars exhaust their thermonuclear fuel and die out. Black holes evaporate. The  density of matter and  the temperature approach zero. The universe becomes cold, dark and inhospitable.  

\end{itemize}

When in this history would one expect to find \igs? Not in the quantum gravity epoch where there isn't even classical spacetime to define a notion of localized subsystem. Not in the early universe.  That period has regularity in the homogeneity and isotropy. But it lacks the complexity that would give an evolutionary advantage to an \ig. Not in the late universe for the same reason. Only in the middle universe do we have both regularities for an \igs\ to exploit and enough complexity to make exploiting those regularities fruitful. And that is where we do find them.

\subsection{\igs\ in the Laboratory?}
\label{lab}
Could these ideas for the evolution of \igs\  be testable  the laboratory?   Would it be possible to vary the mix of regularity, randomness, and complexity to see how whether and how simple \igs\ evolve.  Current experiments in the  evolution of bacteria \cite{Lenski03,WK00} give some hope for this 

\section{Beyond the Quasiclassical Realms}
\label{beyondqc}
The quasiclassical realms are not the only decoherent sets of coarse-grained alternative histories predicted by \HP. There are other realms coarse grained by variables other than quasiclassical ones (e.g.  those in \cite{BH99} in a simple example.) Some of these  could be incompatible with our quasiclassical one in the sense that there is no  finer-grained realm of which both are coarse grainings.  The incompatible  realms give {\it complementary} accounts of the universe somewhat like histories position and histories of momentum for a  single particle. Both are necessary for a complete description. Quantum mechanics does not favor one over the other.

Human IGUSes  focus almost exclusively on the histories of a quasiclassical realm.  When we carry out certain experiments such as measuring the spin of the electron we are using histories of spin that are not quasiclassical. But even there, the apparatus is described and the results recorded in quasiclassical alternatives. Why this almost exclusive focus?

This seeming  question is not a really question for the same reason that asking `Why is the universe comprehensible?' is not a question. It  implicitly assumes that we are somehow outside the universe the box (Fig \ref{box}) and might choose different realms to focus on or even measure\footnote{The author made a related mistake in some early papers where he argued that we evolved to make use of the regularities that quasiclassical realms present. But there is no historical evidence for such selection \cite{Sau93}.}. But we are not outside. We are physical subsystems inside the universe, described in quasiclassical terms, adapted to an environment describable in these terms  whose actions and evolution are described by a history of a quasiclassical realm. We are a feature of a certain quasiclassical realms of our universe not independent of them. 

Can realms incompatible with quasiclassical  ones exhibit IGUSes?  This intriguing question cannot be answered at the present  because we lack a sufficiently general characterization of an IGUS.  We might conjecture that realms with a high level of predictability like the quasiclassical ones would exhibit \igs\ because these present enough regularity over time to permit the
generation of models (schemata) with significant predictive power.  Measures of classicality may help check this conjecture (see e.g. \cite{GH95cl}).  

\section{Conclusion}
\label{conclusion}
It was Everett's insight \cite{Eve57} that  observers  are physical systems within the universe not somehow on the outside observing it.  Observers are describable in quantum mechanical terms as \igs\, subject to quantum mechanical laws (e.g \HP)\,  but play no preferred role in the formulation of quantum mechanics. This change from an `outside' view of observers to an `inside' one has significant consequences for physics. We mention a few examples:

Familiar \igs\ are small subsystems of the universe. Their presence or absence therefore makes only a little change  to the  probabilities supplied by \HP\ for the large scale history of the universe.  But \igs\ are central for  probabilities that predict for the results of {\it our} observations of the universe. That is because probabilities for observations are probabilities supplied by \HP\  {\it conditioned} on data $D$ that describe our observational situation including a description of ourselves. 

Observations of the temperature of the CMB provide a  simple example. What does the  theory \HP\ predict  for the probabilities for our observations of the CMB temperature?  This seeming question is meaningless.  The theory \HP\ predicts probabilities for a four-dimensional time history of the temperature. To predict what we observe these probabilities must be conditioned on data that specify the time from the big bang that we make these observations. 

The `inside' view of observers has significant consequences for the prediction of observations as we have described elsewhere \cite{HH15}. We mention just two examples:  (1) What is called `anthropic selection' is automatic in quantum cosmology --- not a principle, not an option, and not subjective. (2)  The most probable histories to occur are not necessarily the most probable histories to be observed. In models these two consequences significantly affect the probabilities predicting observations of the amount of inflation, the value of the cosmological constant, and the spectrum of the CMB e.g.  \cite{HHH08a,HHH08b,HHH10b,HH13,Her13}

What emerges from the discussion in this paper is that not only is the answer to certain questions affected by the inside view but also the questions themselves. Questions that would seem to be unanswerable when the observer is outside the system become answerable when the observer is seen as an IGUS inside. 

 We provided two examples: First there is Einstein's question about why the universe is comprehensible. We interpreted this as why the histories of the universe exhibit any understandable regularities as all. Our answer is that there must be some regularities through which the universe can be comprehended  otherwise there would be no \igs . Second, we considered the question of why we focus on the quasiclassical realm. The answer becomes a triviality if we understand that we are IGUSes that are  a feature  of a quasiclassical realm and not outside or separate from it. 

We have probably only begun to appreciate the consequences of Everett's insight. 

\acknowledgements

The author thanks Murray Gell-Mann, Thomas Hertog, and Mark Srednicki  for discussions on the quantum mechanics of the universe over many years. He thanks David Krakauer for discussions on experimental evolution. He thanks the Santa Fe Institute for supporting many productive visits there. He thanks FQXI for the invitation to speak at its meeting. This work was supported in part by the National Science Foundation under grant PHY15-04541.

 \appendix
 
 \section{Histories and Realms}
 \label{histories-realms}
 This appendix presents a bare bones description of how the theory \HP\  predicts probabilities for which of a decoherent  set of alternative coarse-grained histories happens in the universe. Many more details and specific models can be found in \cite{classicDH,Har93a}. 
 
 The simplest notion of a set of histories is described by giving a sequence of  yes-no alternatives at a series of times. 
For example: Is a particle in this region $R$ of the box at this time --- yes or no?  This alternative is {\it coarse-grained}  because it does not  follow the particle's position exactly but only whether it is in $R$ or not. The alternative `yes' is represented by the projection operator $P_R$ on the region $R$ amd `no' by $I-P_R$.  More generally coarse-grained yes-no  alternatives at one moment of time are described by an exhaustive set of exclusive Heisenberg picture projection operators $\{P_\alpha(t)\}$, $\alpha=1.2.3.\cdots$ acting in $\Hc$.   These satify:
\be
\label{projections1} 
P_\alpha(t)P_{\alpha'}(t) = \delta_{\alpha\alpha'} P_\alpha(t), \quad \sum_\alpha P_\alpha(t) = I . 
\ee

Projections on bigger subspaces are more coarse grained, projectors on smaller subspaces are finer grained. 

Projection operators representing the same quantity at different times are connected by unitary evolution by the Hamiltonian $H$
\be
P_\alpha(t') = e^{iH(t'-t)}P_\alpha(t) e^{-iH(t'-t)} . 
\label{un-evol}
\ee
For example, for the quasiclassical realm of the model box universe described in Section \ref{qcrealms} the projections would be products of projections  for each subvolume onto ranges of values of the quasiclassical variables  --- averages of energy, momentum, and number. 

A set of alternative coarse-grained histories is specified by a sequence of such sets of orthogonal projection operators at a series of times $t_1,t_2, \cdots t_n$. An individual history corresponds to a particular sequence of events  $\alpha \equiv (\alpha_1,\alpha_2, \cdots, \alpha_n)$ and is represented by the corresponding chain of projections:
\be
C_\alpha \equiv P^n_{\alpha_n}(t_n) \dots P^1_{\alpha_1}(t_1) .
\label{class-closed}
\ee
%An immediate consequence of this and \eqref{projections1} is that 
%\be
%\sum_\alpha C_\alpha =I \ ,
%\label{exhaustC}
%\ee
%showing that the set of histories is exhaustive. 
Branch state vectors corresponding to the individual histories can be defined by
\be
\label{branch}
|\Psi_\alpha\rangle \equiv  C_\alpha |\Psi\rangle . 
\ee

The set of histories (medium) decoheres when the quantum interference between branches is negligible
\be
\label{decoherence}
\langle\Psi_\alpha|\Psi_\beta\rangle \propto \delta_{\alpha\beta}.
\ee
As a consequence of decoherence probabilities $p(\alpha)$  consistent with usual probability rules can be assigned to the individual histories that are 
\be
\label{probs}
p(\alpha) = |||\Psi_\alpha\rangle||^2 = ||C_\alpha |\Psi\rangle||^2.
\ee
A decoherent set of alternative coarse-grained histories is called a `realm' for short.  

This description of histories may seem similar to those for sequences of ideal measurements in Copenhagen quantum mechanics  \cite{Gro52}.  However, there are at least three crucial differences. First, there is no posited separate classical world as in the Copenhagen formulation. It's all quantum. Second, the alternatives represented by the $P$'s are not restricted to measurement outcomes. They might, for example, refer to the orbit to the Moon when no one is looking at it, or to the magnitude of density  fluctuations in the early universe when there were neither observers nor apparatus to measure them. Laboratory measurements can of course be described in terms of correlations between two particular kinds of  subsystems of the universe --- one being measured, the other doing the measuring. But laboratory measurements play no central role in formulating DH, and are just a small part of what it can predict\footnote{Indeed, the Copenhagen quantum mechanics of measured subsystems is an approximation appropriate for measurement situations to the more general quantum mechanics of closed systems. See, e.g. \cite{Har91a} Section II.10.}. Third, the $P$'s can refer to  alternatives describing observers and other IGUSes and to histories of how they formed and what they are doing:  for instance whether they are making measurements or not. 

%%%%%%%%%%%%%%%%%%%%%%%%%%%%%%%%%%%%%%%%%%%%%%%%%%%%%%%

\end{document}